\documentclass[aps,preprint,showpacs,nofootinbib,12pt]{revtex4}
\usepackage{graphicx}
\usepackage{epsfig}
\usepackage[dvips]{color}

\begin{document}

\title{Fraction of the gluonium component in $\eta'$  and $\eta$}

\author{ Hong-Wei Ke$^{1}$   \footnote{khw020056@hotmail.com},
         Xu-Hao~Yuan $^{2}$   \footnote{segoat@mail.nankai.edu.cn} and
         Xue-Qian Li $^{2}$  \footnote{lixq@nankai.edu.cn}
       }

\affiliation{
  $^{1}$ School of Science, Tianjin University, Tianjin 300072, China \\
  $^{2}$ School of Physics, Nankai University, Tianjin 300071, China
  }

\begin{abstract}
\noindent It is interesting to determine the fraction of the
gluonium component in $\eta$ and $\eta'$ which has been under
serious discussion for many years. Measurements on different decay
and/or production modes were employed in literatures, thus larger
uncertainties were unavoidable. In this paper we suggest to
determine the mixing angles of $\eta-\eta'-G$ using the data of
semileptonic decays of $D$ and $D_s$. We extract the mixing angles
$\phi'$, $\phi_G$ and the model parameters simultaneously. Thanks
to the new measurements carried out by CLEO Collaboration and
there are sufficient decay modes to determine both the model
parameters and mixing angles. The mixing angles from data are
$\phi'=(41.5\pm2.0)^\circ$ and $\sin\phi_G=0.00\pm0.36$. Even
though the central value of $\sin\phi_G$ is still zero, an upper
bound is set. Moreover, as suggested in literature, $\eta(1405)$ is a glueball candidate whereas in our
picture, $\eta(1405)$ may be identified as $G$ with glueonium being its main content. Using all the model-parameters obtained
above, we  estimate the
branching ratios of $D_s^+(D^+)\rightarrow
Ge^+\nu_e$ where G is identified as $\eta(1405)$.
\end{abstract}


\maketitle

\section{introduction}
For a long time the mixing of  $\eta-\eta'$ is of a great
theoretical interest because it concerns many aspects of the
underlying dynamic and the structure of pseudoscalar meson. One
can investigate this mixing from two distinct schemes, namely they
may be a mixture  of the octet and singlet of the flavor SU(3), or
$\frac{\sqrt{2}}{2}[\bar u u+\bar dd]$ and $\bar s s$ that are the
mass eigenstates if we can assume an u-d degeneracy. The two
schemes reflect different understandings of the essential physics.
Moreover, the general principle of QCD implies that a gluonic
degree of freedom of $0^{-+}$ may be involved in the physical
states $\eta$ and
$\eta'$\cite{Rosner,Ball:1995zv,Benayoun:1999fv,Benayoun:1999fv2,Benayoun:1999fv3,Cheng2009,Ambrosino2007,
Mathieu:2009sg,Ambrosino:2009sc,Escribano:2007cd,Thomas:2007uy}.
The QCD anomaly indeed may induce such a mixing. It would be
crucially interesting to make a definite judgement if there is a
sizable gluonic component in $\eta$ and $\eta'$ from either an
underlying theory or phenomenology. Obviously even though the QCD
anomaly is well formulated, it is hard to find its contribution to
physical eigenstates because the matrix elements of the anomaly
operator are dominated by the non-perturbative QCD and not
reliably calculable so far. Therefore, one should turn to the
phenomenological studies from which one may gain much information
about the non-perturbative QCD, i.e. the mysterious aspect of the
successful theory. In this work, we are going to determine the
fraction of the glueball component in $\eta$ and $\eta'$ along the
line.

Some researches\cite{Gershtein:1976mv,Chao:1989yp,Degrande:2009ps}
were done on this topic. Generally the authors obtained the mixing
from different decay or production processes
\cite{Feldmann:1998vh}. It can be conjectured that different modes
are governed by different physics mechanisms besides the mixing
which is the goal of our investigation, the uncertainties are not
controllable, therefore the errors would be large. Instead, we
suggest to use the data of semileptonic decays of $D$ and $D_s$
where the reaction mechanism should be the unique. Thus one can
expect that in such a way, uncertainty  could be greatly reduced.
In the theoretical calculations, we employ the light-front-quark
model (LFQM) which has been widely used for calculating the
hadronic matrix elements. In the model, there are two independent
model parameters (see the text for details) which are related to
non-perturbative QCD and should be fixed by fitting data.

In our previous work \cite{Ke:2009mn}, with the data of three
independent measurements on $\mathcal{BR}(D^+\rightarrow \eta
e^+\nu_e)$\cite{Mitchell:2008kb}, $\mathcal{BR}(D_s\rightarrow\eta
e^+ \nu_e)$ and $\mathcal{BR}(D_s\rightarrow\eta' e^+
\nu_e)$\cite{Yelton:2009cm} we determined the $\eta-\eta'$ mixing
$\phi=(39.9\pm2.6(exp)\pm2.3(the))^\circ$ and  predicted the
branching ratio $\mathcal{BR}(D^+\rightarrow
\eta'e^+\nu_e)=(2.12\pm0.23(exp)\pm0.20(the))\times
10^{-4}$\cite{Ke:2009mn} which is consistent with the current data
$\mathcal{BR}^{exp}(D^+\rightarrow
\eta'e^+\nu_e)=(2.16\pm0.53\pm0.07)\times 10^{-4}$ measured by
CLEO collaborator recently\cite{:2010js}. It is noted that there
are three free parameters if the gluonic degree of freedom in
$\eta$ and $\eta'$ is not accounted, namely two are the model
parameters and another is the mixing angle. Instead, if one needs
to consider the extra mixing between quark states with gluonic
degree of freedom, there are four free parameters and at least
four independent measurements are necessary.

Even though in the framework the assumption of null component of
gluonium state is adopted and the predicted branching ratio of
$\mathcal{BR}(D^+\rightarrow \eta'e^+\nu_e)$ is consistent with
data, one still cannot confirm that there indeed is no gluonium
component in $\eta$ and $\eta'$ because the experimental errors
are relatively large. Considering the present data, one can
conclude that the central value of the fraction of gluonium  in
$\eta$ and $\eta'$ is consistent with zero, but while taking into
account the error tolerance, instead, we would only able to obtain
an upper bound of its fraction. Fortunately, thanks to the new
measurements carried out by the CLEO collaboration\cite{:2010js},
the data for all the four decay modes are available, which enable
us to simultaneously fit the two model parameters and  the two
mixing angles. Thus we may determine the fraction of the gluonium
component in $\eta$ and $\eta'$ ( concretely the upper bound of
the gluonium component).

After this introduction, we  describe the working framework and
provide all necessary formulas. In the following section, we present
our numerical results and the last section is devoted to the
discussions and conclusion.

\section{The model calculation}

In the theoretical framework  where gluonium component is not
accounted, the mixing matrix of $\eta-\eta'$ is set as
\cite{kekez,phi,Fajfer:2004mv,Feldmann:1998vh}
\begin{eqnarray}\label{mixing11}
 \left ( \begin{array}{ccc} \eta\\  \eta' \end{array} \right )=
 \left ( \begin{array}{ccc}
  \rm \cos\phi & \rm -\sin\phi \\
  \rm \sin\phi & \rm \cos\phi \end{array} \right )
 \left ( \begin{array}{ccc} \eta_{q} \\  \eta_{s} \end{array}\right ),
 \end{eqnarray}
where $\eta_{q}={1\over\sqrt 2}(u\bar u+d\bar d)$ and
$\eta_{s}=s\bar s$ and it is noted that both of them are not SU(3) singlet. We referred this mixing as scenario-I in this work.

Since the QCD anomaly causes
the mixing between $\eta$ and $\eta'$, there is no any rule to forbid a mixing between
the quark states and a glueball state of the quantum number $0^{-+}$.

As one extends the picture to involve a gluonium component, a new
scenario which we refer as  the scenario-II, was suggested in
Refs.
\cite{Rosner,Ball:1995zv,Benayoun:1999fv,Benayoun:1999fv2,Benayoun:1999fv3,Cheng2009,Ambrosino2007,Mathieu:2009sg,Ambrosino:2009sc}
as
\begin{eqnarray}\label{mixing12}
 \left ( \begin{array}{ccc} \eta\\  \eta' \\G\end{array} \right )=
 \left ( \begin{array}{ccc}
  \rm \cos\phi' & -\rm \sin\phi' & 0\\
  \rm \sin\phi' \cos\phi_G& \rm \cos\phi' \cos\phi_G&\sin\phi_G\\
  -\rm \sin\phi' \sin\phi_G& -\rm \cos\phi' \sin\phi_G&\cos\phi_G\end{array} \right )
 \left ( \begin{array}{ccc} \eta_{q} \\  \eta_{s} \\g\end{array}\right ),
 \end{eqnarray}
where $|g\rangle=|\rm gluonium\rangle$ is a pure gluonium state
and the physical state $G$ was identified as
$\eta(1405)$\cite{Cheng2009}.

As discussed in the introduction, once there are data
on $D^+\rightarrow \eta e^+\nu_e$ available in addition to the other three modes, we will be able to extract $\phi_G$
directly. Moreover new $\mathcal{BR}(D^+\rightarrow \eta
e^+\nu_e)=(11.4\pm0.9\pm0.4)\times 10^{-4}$  deviates from the old datum
and the combined branching ratios of  $\mathcal{BR}(D_s\rightarrow \eta(\eta')
e^+\nu_e)$ \cite{PDG10} demands a
redetermination of the mixing angles in (\ref{mixing12}).

Our strategy is that the mixing angles $\phi'$, $\phi_G$ and the model
parameters $\beta^q_{\eta(\eta')}$, $\beta^s_{\eta(\eta')}$ in the
wave functions of $\eta$ and $\eta'$ are simultaneously determined by fitting solely one
type of data i.e. we theoretically calculate the branching ratios of
$D^+\rightarrow \eta e^+\nu_e$, $D^+\rightarrow \eta' e^+\nu_e$,
$D_s\rightarrow \eta e^+\nu_e$ and $D_s\rightarrow \eta' e^+\nu_e$
and match them with data. Thus we ``extract" the
mixing angles directly from the semileptonic decays of
$D$ and $D_s$.

\begin{figure}
\begin{center}
\begin{tabular}{cc}
\includegraphics[width=9cm]{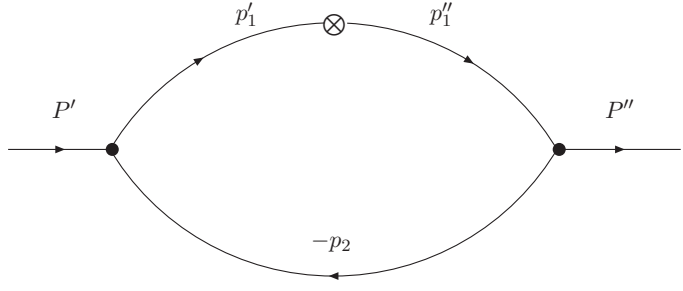}
\end{tabular}
\end{center}
\caption{ Feynman diagram for meson transition amplitude}
\label{fig:LFQM}
\end{figure}

We  use the light front quark model (LFQM) to evaluate the
hadronic transition matrix elements
\cite{Jaus:1999zv,Cheng:2003sm,Hwang:2006cua,Hwang:2006cua2,Hwang:2006cua3,Hwang:2006cua4,Ke:2009ed,Ke:2009ed2,Ke:2009ed3,Li:2010bb,Li:2010bb2},
then obtain the decay widths of $D^+\rightarrow \eta(\eta') e^+
\nu_e$ and $D_s\rightarrow \eta(\eta') e^+ \nu_e$ which are
functions of $\phi'$, $\phi_G$ $\beta^q_{\eta(\eta')}$ and
$\beta^s_{\eta(\eta')}$.  The transition diagram is shown in Fig.
\ref{fig:LFQM}.

The hadronic matrix elements  for $P\rightarrow P$ transition can
be parameterized  as
\begin{eqnarray}\label{2s1}
\langle P(P'')|A_\mu|P(P')\rangle=f_+(q^2){\mathcal{
P}}_\mu+f_-(q^2)q_\mu.
\end{eqnarray}
where $P$ represents the pseudoscalar meson, $P'(P'')$ is the momentum
of initial (final) meson, $q=P'-P''$ and ${ \mathcal{P}}=P'-P''$.

Functions $f_{\pm}(q^2)$ can be calculated in the LFQM and their
explicit expressions were presented as \cite{Cheng:2003sm},
\begin{eqnarray}\label{2s4}
f_+(q^2)=&&\frac{N_c}{16\pi^3}\int dx_2d^2p'_\perp
 \frac{h'_ph''_p}{x_2\hat{N}'_1\hat{N}''_1}
 \left[-x_1(M'^2_0+M''^2_0)-x_2q^2+x_2(m_1'-m_1''^2)\right.\nonumber\\&&
 \left.+x_1(m'_1-m_2)^2+x_1(-m''_1+m_2)^2\right],\nonumber\\
f_-(q^2)=&&\frac{N_c}{16\pi^3}\int dx_2d^2p'_\perp
\frac{2h'_ph''_p}{x_2\hat{N}'_1\hat{N}''_1}
 \left\{x_1x_2M'^2+p'_\perp+m_1'm_2+(-m_1''+m_2)(x_2m_1'+x_1m_2)\right.\nonumber\\&&
 -2\frac{q\cdot \mathcal{P}}{q^2}\left(p'^2_\perp+2\frac{(p'_\perp\cdot
 q_\perp)^2}{q^2}\right) -2\frac{(p'_\perp\cdot
 q_\perp)^2}{q^2}+\frac{(p'_\perp\cdot
 q_\perp)}{q^2}\left[M''^2-x_2(q^2+q\cdot \mathcal{P})\right.\nonumber\\
 &&-(x_2-x_1)M'^2 \left.\left.
 +2x_1M'^2_0-2(m'_1-m_2)(m'_1+m''_1)\right]\right\}.
\end{eqnarray}
where $m'_1,\; m''_1$ and $m_2$ are the corresponding quark
masses, $M'$ and $M''$ are the masses of the initial and final
mesons respectively. All other notations can be found in the
Appendix.

Since the calculations are done in space-like region one need to
analytically continue them to the time-like region. It is
convenient to redefine the matrix elements as
\begin{eqnarray}\label{2s2}
\langle P(P'')|A_\mu|P(P')\rangle=\left({
\mathcal{P}}_\mu-\frac{M'^2-M''^2}{q^2}q_\mu\right)
F_1(q^2)+\frac{M'^2-M''^2}{q^2}q_\mu F_0(q^2).
\end{eqnarray}
The relations among
the form factors are
\begin{eqnarray}
F_1(q^2)=f_+(q^2),\,\, F_0(q^2)=f_+(q^2)+\frac{q^2}{q\cdot{
\mathcal{P}}}f_-(q^2).
\end{eqnarray}

The form factors in the space-like region are parameterized in
a three-parameter form as
 \begin{eqnarray}\label{s14}
 F(q^2)=\frac{F(0)}{1-a\left(\frac{q^2}{M^2}\right)
  +b\left(\frac{q^2}{M^2}\right)^2}.
 \end{eqnarray}
where $F$ represents the form factor $F_{1}$ and $F_{0}$. The
parameters $a,~b$ and $F(0)$ are fixed by performing a
three-parameter fit to the form factors in the space-like region.
We then use these parameters to determine the physical form
factors in the time-like region.

The input quark masses are directly taken from Ref.
\cite{Cheng:2003sm} as $m_u=0.26$ GeV, $m_s=0.37$ GeV, $m_c=1.4$
GeV.

We first need to determine the model parameters for $D$ and $D_s$.
They can be fully determined by fitting their decay constants
which are related to their total widths.
For a pseudoscalar meson the decay constant can be evaluated
\begin{eqnarray}\label{bpp1}
f_P=\frac{\sqrt{N_c}}{16\pi^3}\int
dx_2d^2p'_\perp\frac{\varphi'}{\sqrt{2x_1x_2}\tilde
{M_0'}}4(m_1'x_2+m_2x_1).
 \end{eqnarray}

The decay constants are experimentally measured as $f_{D}^{\rm
exp}=0.221~{\rm MeV}, ~f_{D_s}^{\rm exp}=0.27~ {\rm MeV}$. With
them the model parameters for $\beta_{D_s}$ and $\beta_D$ are
fixed to be $\beta_{D}=0.499$ GeV, $\beta_{D_s}=0.592$ GeV
\cite{Wei:2009nc}. Since the total widths are better measured, the
model parameters of the heavy mesons $D$ and $D_s$ are determined
with higher accuracy.

Once the model parameters at the initial side are fixed, we
would turn to concern the decay products.  At first look, there are six
free parameters $\beta^q_\eta$, $\beta^q_{\eta'}$, $\beta^s_\eta$,
$\beta^s_{\eta'}$, $\phi'$ and $\phi_G$ to be fixed. It seems that
there are not enough equations to determine all these parameters.
However,  as discussed in Ref.\cite{Ke:2009mn} two relations
$\beta^q_\eta=\beta^q_{\eta'}$ and $\beta^s_\eta=\beta^s_{\eta'}$
hold, thus the number of unknowns reduces to four. Matching the theoretically calculated branching ratios with the data
for the four decay modes, we obtain all the values of $\phi'$,
$\phi_G$, $\beta_{\eta(\eta')}^q$, $\beta_{\eta(\eta')}^s$. The
mixing angles are fitted to be $\phi'=(41.5\pm2.0)^\circ$ and
$\sin\phi_G=0.00\pm0.36$($\sin^2\phi_G=0.00\sim0.13$) where the
errors are from the experimental side. The model parameters  are
$\beta_{\eta(\eta')}^q=0.35$ GeV and $\beta_{\eta(\eta')}^s=0.59$
GeV.

With these parameters, assuming $\eta(1405)$ to be the physical
state $G$ whose main content is glueonium $g$, we are able to
estimate the branching ratios of $D_s$ decaying into eta(1405) via
its $q\bar{q}$ components and we obtain
$\mathcal{BR}(D_s^+\rightarrow Ge^+\nu_e)=0\sim 8.6\times 10^{-4}$
and $\mathcal{BR}(D^+\rightarrow Ge^+\nu_e)=0\sim1.1\times
10^{-5}$. Accurate measurements on these semi-leptonic decay modes
may  tell us if the main content of $\eta(1405)$ is glueonium
(i.e. a glueball candidate) and further test the Scenario II for
the mixing.
\section{Discussion}

In Ref.\cite{Ambrosino:2009sc} by fitting the data of several
radiative decay modes such as $\omega\rightarrow\eta\gamma$,
$\rho\rightarrow\eta\gamma$  and $\omega\rightarrow\pi^0\gamma$
${\rm sin^2} \phi_G=0.115\pm 0.036 $\footnote{the error comes from the
$\chi^2$ method.} and $\phi'=(40.4\pm0.6)^\circ$ was fixed if one
sets $ {\rm sin^2}\phi_G$ as a free parameter.

The value of
$\phi'$ obtained from different kinds of experiments are consistent with
each other in a reasonable error tolerance but the central value
of $ {\rm sin^2}\phi_G$ diverges over a
range. The central value of  $ {\rm sin^2}\phi_G$ obtained from the
semileptonic decays of $D$ and $D_s$ inclines  to that the fraction of the gluonium
state in $\eta$ and  $\eta'$ is consistent with zero, but since there exists a relatively large uncertainty
in the data, it is hard to conclude that it does not exist, i.e.
${\rm sin^2} \phi_G=0$. It
is urgent to improve the precision of measurement on these channels
especially on $D(D_s)\rightarrow\eta'e\nu_v$.

In this work, we employ the LFQM model to evaluate the hadronic
transition matrix elements. We fix the model parameters
by fitting data, so that some theoretical uncertainties are involved in those parameters.
Definitely the results are still model dependent because we employ a concrete model: the LFQM. But
since all inputs are taken from the same source (i.e. the data of the semi-leptonic decays of $D$ and $D_s$),
one can expect that relative errors would be partly compensated, so that the
model-dependence of the results is somehow alleviated.

Because of absence of the final state interactions, the
semileptonic decays have obvious advantages for determining the
properties of the produced light hadrons, such as the structure of
$f_0(980)$, $\eta-\eta'$ mixing and even a mixing of pseudoscalar
mesons with glueball.

Moreover, as suggested, assuming that the physical state
$\eta(1405)$ is the mixture $G$ whose main content is glueonium,
we calculate the branching ratios of
$\mathcal{BR}(D_s^+\rightarrow Ge^+\nu_e)$ and and
$\mathcal{BR}(D^+\rightarrow Ge^+\nu_e)$. The semileptonic decay
modes will be measured by the future experiments. Then the
proposed scenario II for the mixing of $q\bar q$, $s\bar s$ and
glueoinum $g$ will be tested.

\section*{Acknowledgments}

This project is supported by the National Natural Science
Foundation of China (NSFC) under Contracts No. 10775073, No.
11075079 and No. 11005079; the Special Grant for the Ph.D. program
of Ministry of Eduction of P.R. China No. 20070055037 and No.
20100032120065.

\appendix

\section{}

Here we list some variables appearing in the context.  The
incoming (outgoing) meson in Fig. \ref{fig:LFQM} has the momentum
${P'}^({''}^)={p_1'}^({''}^)+p_2$ where ${p_1'}^({''}^)$ and $p_2$
are the momenta of the off-shell quark and antiquark and
\begin{eqnarray}\label{app1}
&& p_1'^+=x_1P'^+, \qquad ~~~~~~p_2^+=x_2P'^+, \nonumber\\
&& p'_{1\perp}=x_1P'_{\perp}+p'_\perp, \qquad
 p_{2\perp}=x_2P'_{\perp}-p'_\perp,
 \end{eqnarray}
with  $x_i$ and $p'_\perp$ are internal variables and $x_1+x_2=1$.

The variables $M_0'$, $\tilde {M_0'}$, $h_p'$ and $\hat{N_1'}$ are
defined as
\begin{eqnarray}\label{app2}
&&M_0'^2=\frac{p'^2_\perp+m'^2_1}{x_1}+\frac{p'^2_\perp+m^2_2}{x_2},\nonumber\\&&
\tilde {M_0'}=\sqrt{M_0'^2-(m_1'-m_2)^2}.
 \end{eqnarray}

\begin{eqnarray}\label{app3}
h_p'=(M'^2-M'^2_0)\sqrt{\frac{x_1x_2}{N_c}}\frac{1}{\sqrt{2}\tilde{M_0'}}\varphi',
 \end{eqnarray}
where
\begin{eqnarray}\label{app4}
\varphi'=4(\frac{\pi}{\beta'^2})^{3/4}\sqrt{\frac{dp_z'}{dx_2}}{\rm
 exp}(-\frac{p'^2_z+p'^2_\perp}{2\beta'^2}),
 \end{eqnarray}
with $p_z'=\frac{x_2M_0'}{2}-\frac{m_2^2+p'^2_\perp}{2x_2M_0'}$.
\begin{eqnarray}\label{app5}
\hat{N_1'}=x_1(M'^2-M'^2_0).
 \end{eqnarray}


\begin{thebibliography}{99}
\bibitem{Rosner}
 J. Rosner, Phys. Rev. {\bf D 27}, 1101 (1983);
\bibitem{Ball:1995zv}
  P.~Ball, J.~M.~Frere and M.~Tytgat,
  Phys.\ Lett.\  B {\bf 365}, 367 (1996)
  [arXiv:hep-ph/9508359].
\bibitem{Benayoun:1999fv}
  M.~Benayoun, L.~DelBuono, S.~Eidelman, V.~N.~Ivanchenko and H.~B.~O'Connell,
  Phys.\ Rev.\  D {\bf 59}, 114027 (1999) [arXiv:hep-ph/9902326].
  \bibitem{Benayoun:1999fv2}
 F. J. Gilman and R. Kauffman, Phys. Rev. D {\bf 36}, 2761 (1987).
 \bibitem{Benayoun:1999fv3}
 A. Bramon, R. Escribano and M. D. Scadron, Phys. Lett. B {\bf 503}, 271 (2001).

\bibitem{Cheng2009}
H. Y. Cheng, H. N. Li and K. F. Liu, Phys. Rev. D {\bf 79}, 014024
(2009).



\bibitem{Ambrosino2007}
 F. Ambrosino $et\, al$. [KLOE Collaboration], Phys. Lett. B 648, 267(2007).

\bibitem{Mathieu:2009sg}
  V.~Mathieu and V.~Vento,
  arXiv:0910.0212 [hep-ph].



\bibitem{Ambrosino:2009sc}
  F.~Ambrosino {\it et al.},
  JHEP {\bf 0907}, 105 (2009)
  [arXiv:0906.3819 [hep-ph]].

\bibitem{Thomas:2007uy}
  C.~E.~Thomas,
  JHEP {\bf 0710}, 026 (2007)
  [arXiv:0705.1500 [hep-ph]].

\bibitem{Escribano:2007cd}
  R.~Escribano and J.~Nadal,
  JHEP {\bf 0705}, 006 (2007)
  [arXiv:hep-ph/0703187].




\bibitem{Gershtein:1976mv}
  S.~S.~Gershtein and M.~Y.~Khlopov,
  JETP Lett.\  {\bf 23}, 338 (1976).
\bibitem{Chao:1989yp}
  K.~T.~Chao,
  Nucl.\ Phys.\  B {\bf 317},  597 (1989);
  Nucl.\ Phys.\  B {\bf 335}, 101 (1990).
\bibitem{Degrande:2009ps}
  C.~Degrande and J.~M.~Gerard,
  JHEP {\bf 0905}, 043 (2009)
  [arXiv:0901.2860 [hep-ph]].



\bibitem{Feldmann:1998vh}
  T.~Feldmann, P.~Kroll and B.~Stech,
  Phys.\ Rev.\  D {\bf 58}, 114006 (1998)
  [arXiv:hep-ph/9802409].

\bibitem{Ke:2009mn}
  H.~W.~Ke, X.~Q.~Li and Z.~T.~Wei,
  Eur.\ Phys.\ J.\  C {\bf 69}, 133 (2010)
  [arXiv:0912.4094 [hep-ph]].


\bibitem{Mitchell:2008kb}
  R.~E.~Mitchell {\it et al.}  [CLEO Collaboration],
  Phys.\ Rev.\ Lett.\  {\bf 102}, 081801 (2009)
  [arXiv:0802.4222 [hep-ex]].
\bibitem{Yelton:2009cm}
  J.~Yelton {\it et al.}  [CLEO Collaboration],
  arXiv:0903.0601 [hep-ex].

\bibitem{:2010js}
 J. Yelton  {\it et al.}  [CLEO Collaboration],
  arXiv:1011.1195 [hep-ex].


\bibitem{kekez} D. Kekez, D. Klabu$\check{c}$ar and M. Scadron, J.
Phys. {\bf G 27}, 1775 (2001).




\bibitem{phi}
M. Scadron, Phys.\ Rev.\ D {\bf 26}, 239 (1982).
\bibitem{Fajfer:2004mv}
  S.~Fajfer and J.~F.~Kamenik,
  Phys.\ Rev.\  D {\bf 71}, 014020 (2005)
  [arXiv:hep-ph/0412140].









\bibitem{PDG10}
  K.~Nakamura {\it et al.}  [Particle Data Group],
  J.\ Phys.\ G {\bf 37}, 075021 (2010).




\bibitem{Jaus:1999zv}
  W.~Jaus,
  Phys.\ Rev.\  D {\bf 60}, 054026 (1999).

\bibitem{Cheng:2003sm}
  H.~Y.~Cheng, C.~K.~Chua and C.~W.~Hwang,
  Phys.\ Rev.\  D {\bf 69}, 074025 (2004).


\bibitem{Hwang:2006cua}
  C.~W.~Hwang and Z.~T.~Wei,
  J.\ Phys.\ G {\bf 34}, 687 (2007).
%
\bibitem{Hwang:2006cua2}
  C.~D.~Lu, W.~Wang and Z.~T.~Wei,
  Phys.\ Rev.\  D {\bf 76}, 014013 (2007)
  [arXiv:hep-ph/0701265].
%
\bibitem{Hwang:2006cua3}
  H.~W.~Ke, X.~Q.~Li and Z.~T.~Wei,
  Phys.\ Rev.\  D {\bf 77}, 014020 (2008).
%
\bibitem{Hwang:2006cua4}
  Z.~T.~Wei, H.~W.~Ke and X.~Q.~Li,
  Phys.\ Rev.\  D {\bf 80}, 094016 (2009)
  [arXiv:0909.0100 [hep-ph]].


\bibitem{Ke:2009ed}
  H.~W.~Ke, X.~Q.~Li and Z.~T.~Wei,
  Phys.\ Rev.\  D {\bf 80}, 074030 (2009)
  [arXiv:0907.5465 [hep-ph]].
  \bibitem{Ke:2009ed2}
  H.~W.~Ke, X.~Q.~Li, Z.~T.~Wei and X.~Liu,
  Phys.\ Rev.\  D {\bf 82}, 034023 (2010)
  [arXiv:1006.1091 [hep-ph]].
  \bibitem{Ke:2009ed3}
  H.~W.~Ke, X.~Q.~Li and X.~Liu,
  arXiv:1002.1187 [hep-ph].

\bibitem{Li:2010bb}
  G.~Li, F.~l.~Shao and W.~Wang,
  Phys.\ Rev.\  D {\bf 82}, 094031 (2010)
  [arXiv:1008.3696 [hep-ph]].
\bibitem{Li:2010bb2}  P.~Colangelo, F.~De Fazio and W.~Wang,
  Phys.\ Rev.\  D {\bf 81}, 074001 (2010)
  [arXiv:1002.2880 [hep-ph]].





\bibitem{Wei:2009nc}
  Z.~T.~Wei, H.~W.~Ke and X.~F.~Yang,
  Phys.\ Rev.\  D {\bf 80}, 015022 (2009)
  [arXiv:0905.3069 [hep-ph]].











\end{thebibliography}
\end{document}